\documentclass[trackchanges]{aastex701}

\usepackage{amsmath,amssymb,amsfonts}
\usepackage{mathtools}
\usepackage{bm,bbm}
\usepackage{graphicx}
\usepackage{xcolor}
\usepackage{multirow,tabularx}
\usepackage{comment}
\usepackage{aas_macros}
\hypersetup{unicode=true}
\usepackage{amsthm}

\newcommand{\pd}{{\rm d}}

\newcommand{\nhat}{\hat{\vec{n}}}

\newcommand{\bvec}{\vec{\beta}}
\newcommand{\vvec}{\vec{v}}
\newcommand{\cov}{\mathbbm{Cov}}

\begin{document}

\title{A General Formulation of the Kinematic Dipole as a Functional of Selection and Source Properties: \\ Beyond the Ellis--Baldwin Approximation}

\author[orcid={0000-0001-8416-7673},sname='Takeuchi']{Tsutomu T.\ Takeuchi}
\affiliation{Division of Particle and Astrophysical Science, Nagoya University, Japan}
\affiliation{The Research Center for Statistical Machine Learning, the Institute of Statistical Mathematics, Japan}
\email[show]{tsutomu.takeuchi.ttt@gmail.com}  



\begin{abstract}

The dipole anisotropy in galaxy and QSO number counts induced by the observer's motion, known as the kinematic dipole, provides an important test of cosmological isotropy and an {independent comparison} with the Cosmic Microwave Background (CMB) dipole. 
Traditionally, the Ellis and Baldwin expression $\mathcal{A}=2+x(1+\alpha)$ has been widely adopted, assuming power-law number counts and a single power-law spectral energy distribution (SED). 
Realistic surveys, however, involve a range of non-ideal effects, including {heterogeneous source SEDs}, finite instrumental bandpasses, non-power-law number counts, multi-band photometry, photo-$z$ selections, and direction-dependent or stochastic detection limits. 
By explicitly incorporating these factors, we develop a framework that describes the kinematic dipole as a functional of the source population and the selection criteria of the catalog.
We show that the dipole amplitude is not described by a single index, but is instead given by a functional $\mathcal{A}[\mathcal{W},f]$, defined as the Doppler response of the {catalog selection} acting on the underlying source population. 
We demonstrate that the classical Ellis--Baldwin result is recovered as a {controlled limiting case} of this formalism, and clarify the relation between the theoretical coefficient $\mathcal{A}$ and the dipole vector estimated from finite catalogs, thereby separating theoretical response from statistical uncertainty. 
We further show that realistic mixed populations, such as AGN- and star-forming-galaxy radio samples, naturally lead to {population-dependent effective coefficients}. 
This framework facilitates {survey-specific predictions} of the kinematic dipole, for reinterpreting reported discrepancies among existing measurements, and is directly applicable to future wide-area, multi-band surveys.

\end{abstract}



\section{Introduction}

Extragalactic source counts from galaxies and QSOs (quasi-stellar objects) provide one of the most direct probes of large-scale isotropy on the sky \citep[e.g.][]{2012MNRAS.427.1994G,2017MNRAS.464..768B,2021A&A...653A...9S}. 
If the underlying source population is statistically isotropic in the cosmic rest frame, the observer's motion with velocity
\begin{align}
    \bvec \equiv \vvec/c
\end{align}
produces a dipolar modulation in the observed counts.
This kinematic dipole originates from two relativistic effects: aberration, which distorts the apparent solid angle, and Doppler boosting of the observed fluxes, which in turn modifies the survey selection.

The first systematic treatment of this effect was given by \citet{1984MNRAS.206..377E}. 
Assuming that the cumulative counts obey a power law,
\begin{align}
    N(>S)\propto S^{-x},
\end{align}
and that each source follows a single power-law spectrum \(S_\nu\propto \nu^{-\alpha}\), they showed that the dipole amplitude is
\begin{align}
    \mathcal A = 2+x(1+\alpha).
    \label{eq:dipole_amplitude_EB}
\end{align}
This simple expression has remained the standard reference ever since \citep{1984MNRAS.206..377E,2021A&A...653A...9S}.

Subsequent observational work, largely in radio surveys, has focused on how practical issues, such as sky masks, incompleteness, flux calibration, estimator bias, and variance, influence the measured dipole \citep[e.g.][]{2009ApJ...692..887C,2013A&A...555A.117R,2015APh....61....1T,2021A&A...653A...9S,2023A&A...675A..72W,2024MNRAS.535L..49V,2025PhRvD.111l3547V}. 
Similar analyses now exist at optical and infrared wavelengths, and the CMB dipole itself has been re-examined for possible intrinsic contributions. 
Taken together, these studies show that observed dipoles are generally a mixture of kinematic, large-scale-structure, and survey-systematic effects \citep[e.g.,][]{2012MNRAS.427.1994G}.

\begin{figure}[t]
    \centering
    \includegraphics[width=\linewidth]{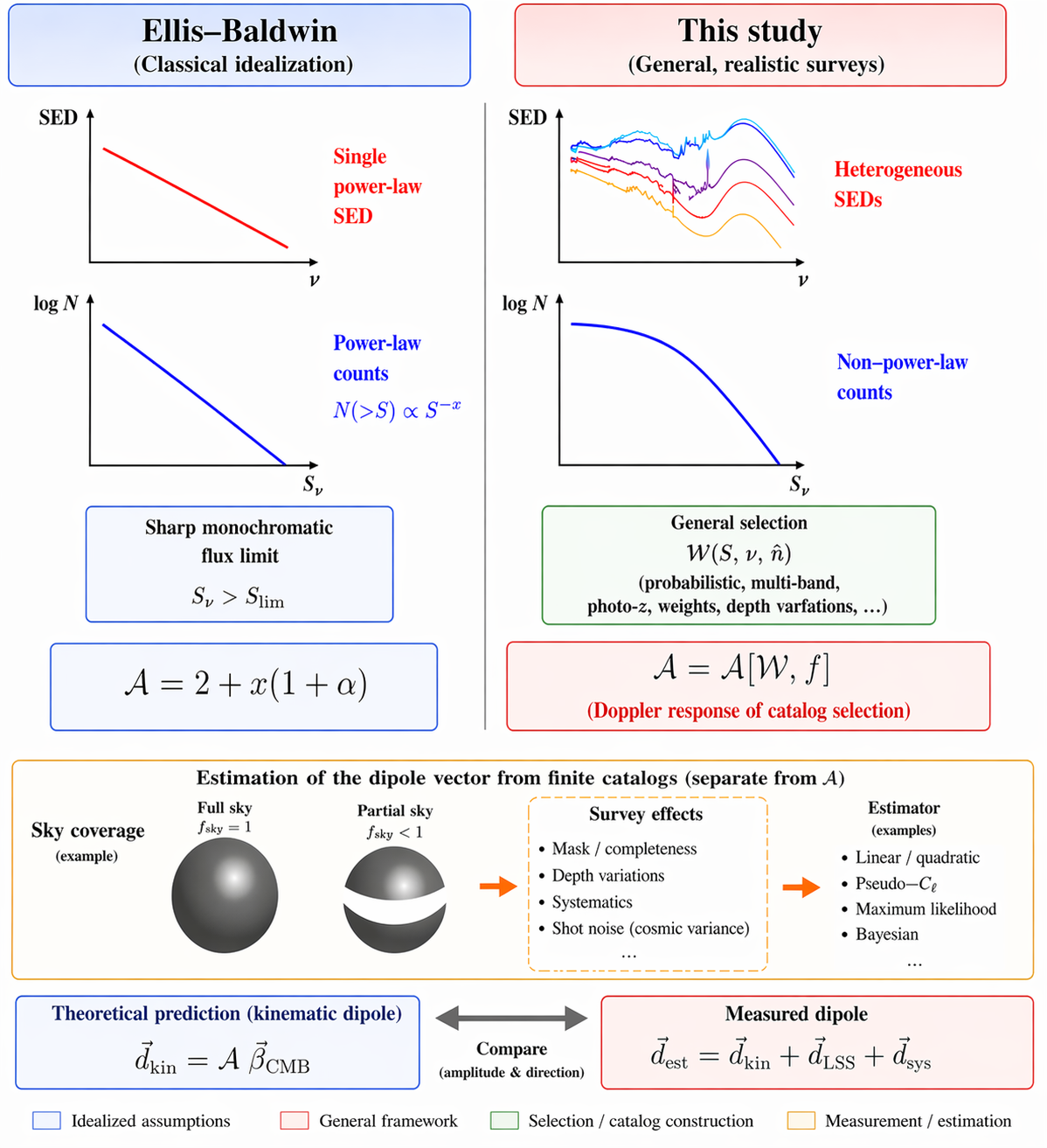}
    \caption{{Schematic comparison between the classical Ellis--Baldwin formulation (left) and the generalized framework developed here (right), together with the distinction between intrinsic kinematic response and dipole estimation from finite catalogs (lower panels). 
    The classical treatment assumes power-law source counts, a single power-law spectral energy distribution (SED), and a sharp monochromatic flux limit, giving $\mathcal A=2+x(1+\alpha)$ \citep{1984MNRAS.206..377E}. 
    The generalized framework allows heterogeneous source populations, non-power-law counts, finite instrumental bandpasses, and multidimensional catalog selections summarized by $\mathcal W(S,\nu,\hat n)$, yielding $\mathcal A=\mathcal A[\mathcal W,f]$. 
    The lower panels emphasize that $\mathcal A$ is determined by the source population $f$ and survey selection $\mathcal W$, whereas incomplete sky coverage, masks, depth variations, noise, and estimator choice affect the recovery of the dipole vector from finite data rather than $\mathcal A$ itself. 
    This distinction underlies the comparison between the theoretical prediction $\vec d_{\rm kin}=\mathcal A\,\vec\beta_{\rm CMB}$ and the measured dipole $\vec d_{\rm est}$.}}
    \label{fig:dipole_estimation}
\end{figure}

However, most theoretical predictions still rely on the idealized assumptions behind eq.~\eqref{eq:dipole_amplitude_EB}. 
In practice, number counts rarely follow strict power laws, and spectral energy distributions (SEDs) vary significantly between sources. 
Furthermore, real-world observations must account for finite instrumental bandpasses and selection criteria that are often stochastic or direction-dependent \citep{2021Univ....7..107S}. 
As large, deep, multi-band surveys become routine, it is essential to account for these observational complexities in the theoretical framework. 
Figure~\ref{fig:dipole_estimation} illustrates the difference between the classical approach and the generalized formalism developed here.

A substantial fraction of previous work has focused on the behavior of dipole \emph{estimators}, while the theoretical coefficient itself has often retained simplified assumptions.
In this work, we focus on formulating the kinematic dipole from a general parent population and a general SED, including finite filters, and expressing the observer-motion effect as the response of a multidimensional selection function.
We therefore define the dipole amplitude as a functional of the selection function and source population, rather than a fixed numerical constant.
This construction directly accommodates the complex selection processes of modern wide-area surveys. 
It accounts for color cuts, photometric redshift selections, and direction-dependent survey depths—factors that are now standard in catalogs such as those by, e.g., \citet{2021ApJ...908L..51S} and \citet{2025JKPS...86..145A}.

Throughout this paper, the cosmic rest frame (denoted by a prime) is assumed statistically isotropic and free of an intrinsic dipole.
This defines the reference model for the kinematic response.
In finite catalogs, the measured dipole may additionally contain large-scale-structure and survey-systematic components, and their separation from the kinematic term is discussed in Section~\ref{sec:discussion}.
The CMB dipole provides the most direct estimate of the observer velocity $\bvec_{\rm CMB}$, so the expected source-count dipole is
\begin{align}
    \vec d_{\rm kin} \equiv \mathcal A\,\bvec_{\rm CMB}.
\end{align}
Observed estimates $\vec d_{\rm est}$ may differ from this prediction because of sample variance and observational systematics.

The paper is organized as follows.
Section~\ref{sec:definitions} introduces notation and observables, including band fluxes for general SEDs and filters (Section~\ref{subsec:band_flux}).
{Section~\ref{sec:general_formalism} develops the general selection formalism and derives the first-order kinematic dipole coefficient as the functional response $\mathcal A[\mathcal W,f]$.}
{Section~\ref{sec:practical_response} shows how this response can be evaluated in practice, including flux-response coefficients for general SEDs and filters, finite-catalog implementations, and chain-rule decompositions relevant to flux thresholds, multiband cuts, and photometric-redshift selections.}
{Section~\ref{sec:eb_limit} recovers the Ellis--Baldwin formula as a controlled limiting case, and Section~\ref{sec:mixed_population_application} presents a mixed-population application to radio source counts with AGN and SFG contributions.}
{Section~\ref{sec:discussion} discusses the relation between the theoretical response coefficient and dipoles estimated from data, including estimator methodology, Bayesian extensions, and the interpretation of reported tensions with the CMB dipole.}
{Additional details on estimator properties and statistical uncertainties are given in Appendices~\ref{app:poisson_likelihood} and~\ref{app:fisher_covariance}.}

\section{Definitions of Notation and Observables}
\label{sec:definitions}

\subsection{Notation conventions}

Throughout this paper, three-dimensional vectors in real space, such as velocities, dipole vectors, and directions, are denoted by arrows, $\vec{\phantom{x}}$.
Elements of multidimensional parameter spaces are denoted by bold symbols, $\bm y$.
{A prime ($'$) denotes quantities defined in the cosmic rest frame and is reserved exclusively for that purpose throughout this paper.}

\subsection{Direction and Doppler factor}

Let $\nhat$ denote the direction of observation and $\bvec$ the velocity of the observer.
The Doppler factor is defined by
\begin{align}
	\delta(\nhat)
	\equiv
	\frac{1}{\gamma(1-\bvec\cdot\nhat)},
	\qquad
	\gamma\equiv(1-\beta^2)^{-1/2},
\end{align}
where $\beta\equiv|\bvec|$.
For $\beta\ll1$, the first-order expansion gives
\begin{align}
	\delta(\nhat)
	&=
	1+\bvec\cdot\nhat+\mathcal O(\beta^2),
	\\
	\ln\delta(\nhat)
	&=
	\bvec\cdot\nhat+\mathcal O(\beta^2).
\end{align}
These standard relations underlie essentially all treatments of the kinematic dipole in source counts \citep[e.g.,][]{1984MNRAS.206..377E,2021A&A...653A...9S}.

\subsection{Band fluxes for general SEDs and filters}
\label{subsec:band_flux}

The basic observables considered in this work are the fluxes measured in individual photometric bands, $S_{\rm b}$.
We denote the spectral flux density in the cosmic rest frame by
\begin{align}
	{S'_{\nu'}(\nu';z',\bm\psi)},
\end{align}
{where $z'$ is the cosmological redshift in the cosmic rest frame, and $\bm\psi$ denotes intrinsic source parameters characterizing the spectral energy distribution (SED).}
{The observed redshift may differ from $z'$ through the kinematic motion of the observer.}

For a moving observer viewing the same source in the direction $\nhat$, the observed frequency transforms as
\begin{align}
	\nu=\delta(\nhat)\nu'.
\end{align}

For the galaxy and QSO catalogs considered here, sources are effectively unresolved (point-like) relative to the angular resolution of the survey.
This approximation is standard for wide-area dipole analyses based on radio, infrared, optical, and QSO catalogs, where individual source morphology is not explicitly modeled in the estimator stage \citep{2009ApJ...692..887C,2021Univ....7..107S,2025A&A...697A.112W}.
The relevant observable is therefore the flux density integrated over the apparent source solid angle,
\begin{align}
	S_\nu(\nu)
	\equiv
	\int I_\nu(\nu,\hat{\bm n})\,\pd\Omega.
\end{align}

The Lorentz invariance of $I_\nu/\nu^3$ implies
\begin{align}
	I_\nu(\nu,\hat{\bm n})
	=
	\delta^3
	I'_{\nu'}\!\left(\frac{\nu}{\delta},\hat{\bm n}'\right),
\end{align}
while aberration gives
\begin{align}
	\pd\Omega=\delta^{-2}\pd\Omega'.
\end{align}
{These are standard results of relativistic radiative transfer \citep[e.g.,][]{1979rpa..book.....R,peebles1993principles}.}

Combining them, the flux density of a point source transforms as
\begin{align}
	S_\nu(\nu)
	&=
	\int I_\nu\,\pd\Omega
	=
	\delta\int I'_{\nu'}\,\pd\Omega'
	=
	\delta\,S'_{\nu'}(\nu/\delta).
	\label{eq:Snu_transform_point}
\end{align}
This transformation is the monochromatic basis of the classical Ellis--Baldwin derivation and of later radio dipole analyses \citep{1984MNRAS.206..377E,2015APh....61....1T,2021A&A...653A...9S}.

Let $R_{\rm b}(\nu)$ denote the transmission (response) function of photometric band ${\rm b}$.
The observed band flux is then
\begin{align}
	S_{\rm b}
	\equiv
	\int_0^\infty R_{\rm b}(\nu)\,S_\nu(\nu)\,\pd\nu.
	\label{eq:Sb_def}
\end{align}
Finite instrumental bandpasses are unavoidable in realistic surveys, and broad-band selections are central to modern optical, infrared, and radio source catalogs.
It is therefore natural to formulate the kinematic response directly at the band-flux level rather than only through an idealized monochromatic flux density.

Substituting equation~\eqref{eq:Snu_transform_point} into equation~\eqref{eq:Sb_def} gives
\begin{align}
	S_{\rm b}(\delta)
	&=
	\int_0^\infty
	R_{\rm b}(\nu)\,
	\delta\,S'_{\nu'}\!\left(\nu/\delta;{z'},\bm\psi\right)\,\pd\nu
	\notag\\
	&=
	\delta^2
	\int_0^\infty
	R_{\rm b}(\delta\nu')\,S'_{\nu'}(\nu';{z'},\bm\psi)\,\pd\nu'.
	\label{eq:Sb_transform}
\end{align}
where the second line follows from the change of variables $\nu'=\nu/\delta$.
Equation~\eqref{eq:Sb_transform} makes explicit that the observer motion affects both the source spectrum and the position of the filter response in frequency space.
This effect is negligible only in the narrow-band limit.

All derived quantities constructed from band fluxes, such as colors or photometric-redshift estimates, inherit their dependence on $\delta$ through the transformation of $S_{\rm b}$.
This point is particularly relevant for modern catalogs whose target selection is performed in multi-dimensional photometric space rather than by a single sharp flux threshold.

\section{General selection formalism and the kinematic dipole}
\label{sec:general_formalism}

\subsection{Selection function and parent population}

In practical surveys, catalog membership is rarely determined by a single sharp flux threshold.
Source inclusion generally depends on photometric uncertainties, quality flags, star--galaxy separation, color cuts, photometric-redshift selections, spatially varying depth, masks, incompleteness corrections, and weighting schemes.
Such ingredients are standard in modern source-catalog construction across optical, infrared, and radio surveys, where selection is effectively multidimensional rather than one-dimensional in flux alone \citep{2021Univ....7..107S,2023A&A...675A..72W,2025A&A...697A.112W}.
We represent these effects by a unified selection function
\begin{align}
	\mathcal W(\bm y,\nhat)\in[0,1],
\end{align}
where $\mathcal W=1$ denotes certain inclusion, $\mathcal W=0$ complete exclusion, and intermediate values represent probabilistic selection or weights.

The source descriptor vector is written as
\begin{align}
	\bm y=(z,\bm\psi,S_{{\rm b}_1},S_{{\rm b}_2},\ldots),
\end{align}
where $z$ is redshift, $\bm\psi$ denotes intrinsic parameters describing the spectral energy distribution (SED), and $S_{\rm b}$ are observed band fluxes.
No assumption is made that all sources share a common SED or a single effective spectral index.
Additional derived quantities, such as colors, photometric-redshift summaries, morphology indicators, or classification scores, may be appended without changing the formal structure.

Here $\bm y$ denotes the catalog-level descriptor entering the selection function.
It may contain both intrinsic source labels and observer-frame measured quantities.
The dependence on observer motion is encoded through the mapping
\begin{align}
	\bm y\mapsto\bm y_\delta,
\end{align}
so that all kinematic effects are treated consistently at the level of the selected catalog variables.

We define the isotropic parent population in the cosmic rest frame by
\begin{align}
	f(\bm y)\,\pd\bm y
	\equiv
	\frac{\pd N}{\pd\Omega\,\pd\bm y}\,\pd\bm y,
\end{align}
and assume that it contains no intrinsic dipole component.
This corresponds to the standard interpretation in which anisotropy in the observed source counts is generated by observer motion plus non-kinematic foreground contributions, rather than by a primordial dipole in the parent distribution \citep{2012MNRAS.427.1994G,2021PhRvL.127j1301F}.

Observer motion acts through the transformation of observables,
\begin{align}
	\bm y\mapsto\bm y_\delta,
\end{align}
induced by the Doppler factor $\delta$.
With this convention, the kinematic response enters through the selection function rather than through an explicit redefinition of $f$.

This framework applies to realistic catalogs with heterogeneous SEDs, non-power-law counts, finite filters, and direction-dependent deterministic or stochastic selection.

\subsection{Observed counts under observer motion}

Aberration transforms the solid-angle element as
\begin{align}
	\pd\Omega=\delta^{-2}\pd\Omega'
\end{align}
\citep[e.g.,][]{1979rpa..book.....R,peebles1993principles},
so number counts per observed solid angle acquire the standard prefactor $\delta^2$.
This geometrical factor is common to all source populations and is independent of the detailed catalog construction.

Combining this effect with the mapping $\bm y\mapsto\bm y_\delta$, the observed surface density becomes
\begin{align}
	n(\nhat)
	=
	\delta^2(\nhat)
	\int
	f(\bm y)\,
	\mathcal W(\bm y_\delta,\nhat)\,
	\pd\bm y.
	\label{eq:n_general}
\end{align}
Equation~\eqref{eq:n_general} is the master counting formula of the present formalism.
It states that observer motion affects source counts through two distinct channels: a universal angular Jacobian and a survey-dependent deformation of the selection boundary.

This form also makes contact with practical dipole analyses based on weighted source maps, masked skies, and threshold-dependent subsamples, where the effective counting operator is never purely geometric \citep{2009ApJ...692..887C,2015APh....61....1T,2021A&A...653A...9S}.

\subsection{First-order dipole coefficient}

Expanding equation~\eqref{eq:n_general} to first order in $\beta$, equivalently in $\ln\delta$, gives
\begin{align}
	n(\nhat)
	=
	n_0(\nhat)
	+
	(\bvec\cdot\nhat)
	\left[
	2\,n_0(\nhat)+R(\nhat)
	\right]
	+
	\mathcal O(\beta^2),
\end{align}
where
\begin{align}
	n_0(\nhat)
	&\equiv
	\int
	f(\bm y)\,
	\mathcal W(\bm y,\nhat)\,
	\pd\bm y,
	\\
	R(\nhat)
	&\equiv
	\int
	f(\bm y)
	\left.
	\frac{\pd}{\pd\ln\delta}
	\mathcal W(\bm y_\delta,\nhat)
	\right|_{\delta=1}
	\pd\bm y.
\end{align}
Here $n_0(\nhat)$ is the catalog surface density in the absence of observer motion, and $R(\nhat)$ is the Doppler response of the survey selection.

The fractional fluctuation is therefore
\begin{align}
	\frac{n(\nhat)-n_0(\nhat)}{n_0(\nhat)}
	=
	\mathcal A(\nhat)\,\bvec\cdot\nhat,
\end{align}
with
\begin{align}
	\mathcal A(\nhat)
	=
	2+\frac{R(\nhat)}{n_0(\nhat)}.
	\label{eq:A_general}
\end{align}

The first term is the universal aberration contribution.
The second term is the Doppler response of the source population and catalog selection.
Equation~\eqref{eq:A_general} is the central result of this work: the kinematic dipole coefficient is expressed as a functional of the parent population and the survey selection.

In the special case of a uniform sky and a sharp monochromatic flux threshold, equation~\eqref{eq:A_general} reduces to the familiar Ellis--Baldwin form discussed later.
For realistic surveys, however, $\mathcal A$ can depend on bandpass choice, source mixture, threshold definition, and sky position.

The present analysis isolates the observer-motion contribution.
Other anisotropy sources, including magnification bias, lensing, relativistic projection effects, large-scale structure, and survey systematics, are conceptually separate and may be added as independent corrections when required.
This separation is consistent with recent observational discussions in which the measured dipole is interpreted as a superposition of kinematic and non-kinematic components \citep{2012MNRAS.427.1994G,2025A&A...697A.112W}.

\section{Practical evaluation of the selection response}
\label{sec:practical_response}

\subsection{Flux-response coefficient for general SEDs and filters}
\label{subsec:flux_response}

From equation~\eqref{eq:Sb_transform}, we define
\begin{align}
    I_{\rm b}(\delta)
    \equiv
    \int_0^\infty
    R_{\rm b}(\delta\nu')\,
    S'_{\nu'}(\nu';z',\bm\psi)\,
    \pd\nu',
\end{align}
so that
\begin{align}
    S_{\rm b}(\delta)=\delta^2 I_{\rm b}(\delta).
\end{align}

The first-order logarithmic response at $\delta=1$ is
\begin{align}
    q_{\rm b}(\bm\psi)
    \equiv
    \left.
    \frac{\pd\ln S_{\rm b}}{\pd\ln\delta}
    \right|_{\delta=1}
    =
    2+
    \left.
    \frac{\pd\ln I_{\rm b}}{\pd\ln\delta}
    \right|_{\delta=1},
    \label{eq:qb_def}
\end{align}
which we call the Doppler response coefficient of band ${\rm b}$.
This quantity depends on both the intrinsic SED of the source and the instrumental bandpass.
In the narrow-band limit with a local power-law spectrum, it reduces to the familiar $1+\alpha$ factor of the Ellis--Baldwin approximation.

For broad filters or structured spectra, the effective response can differ significantly from a single spectral-index description.
This matters for modern surveys, whose selections rely on wide photometric bands or multi-band combinations rather than monochromatic fluxes.

Differentiating under the integral sign gives
\begin{align}
    \left.
    \frac{\pd\ln I_{\rm b}}{\pd\ln\delta}
    \right|_{\delta=1}
    =
    \frac{
    \displaystyle
    \int_0^\infty
    \nu'
    {\frac{\pd R_{\rm b}}{\pd\nu}(\nu')}\,
    S'_{\nu'}(\nu';z',\bm\psi)\,
    \pd\nu'
    }{
    \displaystyle
    \int_0^\infty
    R_{\rm b}(\nu')\,
    S'_{\nu'}(\nu';z',\bm\psi)\,
    \pd\nu'
    }.
    \label{eq:filter_term}
\end{align}
Equations~\eqref{eq:qb_def} and~\eqref{eq:filter_term} therefore allow $q_{\rm b}$ to be computed for arbitrary SEDs and filters.

Equation~\eqref{eq:filter_term} shows that the correction to the universal factor $2$ is set by how rapidly the filter transmission changes across frequencies, weighted by the source spectrum.
When the filter is very narrow or nearly flat over the relevant range, the correction approaches the standard monochromatic limit.

\subsection{Finite-catalog implementation}
\label{subsec:discrete_sum}

For a catalog of $N$ objects $\{\bm y_i\}$, the functional coefficient $\mathcal A[\mathcal W,f]$ can be evaluated directly as
\begin{align}
    \mathcal A
    \approx
    2+
    \frac{
    \displaystyle
    \sum_i
    \left.
    \frac{\pd}{\pd\ln\delta}
    \mathcal W((\bm y_i)_\delta,\nhat_i)
    \right|_{\delta=1}
    }{
    \displaystyle
    \sum_i
    \mathcal W(\bm y_i,\nhat_i)
    }.
    \label{eq:A_discrete}
\end{align}

This discrete sum naturally incorporates object-by-object variations in SED, multi-band photometry, photometric-redshift selection, and direction-dependent survey depth.
It also shows that no global effective $(x,\alpha)$ pair is needed once the relevant catalog data are available.

In practice, the sum can be computed from the science catalog itself, from a deeper parent sample, or from simulated mock catalogs with matched selection functions.
Such forward-modeling techniques are already common in large-survey systematics studies and can be readily adapted for dipole-response calibration \citep{2023A&A...675A..72W,2025A&A...697A.112W}.
Recent analyses using hierarchical or survey-aware modeling further demonstrate that realistic selection and systematics treatments are essential for robust cosmological interpretation \citep{2024MNRAS.535L..49V, 2025PhRvD.111l3547V}.

\subsection{Chain-rule decomposition of realistic selections}
\label{subsec:chain_rule_response}

Suppose the selection function depends on a vector of measured quantities $\vec u$ (fluxes, colors, photometric redshifts, morphology indicators, etc.):
\begin{align}
    \mathcal W(\bm y,\nhat)
    =
    \mathcal W(\vec u(\bm y),\nhat).
\end{align}
Observer motion induces the transformation $\vec u\mapsto\vec u_\delta$, so the response of the selection function becomes
\begin{align}
    \left.
    \frac{\pd}{\pd\ln\delta}
    \mathcal W(\bm y_\delta,\nhat)
    \right|_{\delta=1}
    =
    \sum_k
    \frac{\partial\mathcal W}{\partial u_k}
    \left.
    \frac{\pd u_{k,\delta}}{\pd\ln\delta}
    \right|_{\delta=1}.
    \label{eq:chain_rule}
\end{align}

The dipole response is therefore set by two factors: the sharpness of the selection boundary (encoded in the partial derivatives) and the Doppler response of the relevant observables.
This decomposition is particularly convenient because modern catalog pipelines are modular: photometry produces colors, colors feed photo-$z$ estimators, and these feed the final target mask.

Equation~\eqref{eq:chain_rule} also makes clear that a weakly varying observable can still produce a large response if the boundary is sharp, while a strongly shifted observable contributes little if the selection is broad and probabilistic.

\subsection{Flux thresholds, multiband cuts, and photometric-redshift selections}

For a simple deterministic flux threshold,
\begin{align}
    \mathcal W(\bm y)
    =
    \Theta(S_{\rm b}-S_{\min})\,
    \mathcal W_{\rm other}(\bm y),
\end{align}
the dominant contribution comes from sources near the limit:
\begin{align}
    \left.
    \frac{\pd S_{\rm b}}{\pd\ln\delta}
    \right|_{\delta=1}
    =
    q_{\rm b}(\bm\psi)\,S_{\rm b}.
\end{align}
This recovers the usual intuition that dipole sensitivity is concentrated near the threshold, but now with a source-dependent bandpass response.

For realistic multiband catalogs, additional contributions arise from colors, photometric redshift estimators, morphology classifiers, and other derived quantities.
Typical examples include color wedges for AGN selection, photo-$z$ window cuts for tomographic samples, and star--galaxy separation boundaries in optical surveys.
Once the full pipeline is encoded in $\mathcal W(\bm y,\nhat)$, the corresponding dipole coefficient follows directly from the same formalism.

The practical message is simple: increasingly sophisticated catalog pipelines do not invalidate kinematic dipole predictions; they simply redefine the response coefficient that must be used when comparing measured dipoles with the CMB velocity.

\section{Recovery of the Ellis--Baldwin formulation as a controlled limit}
\label{sec:eb_limit}

We now show how the classical Ellis--Baldwin expression emerges from the general formalism through a sequence of controlled idealizations.
The goal is to make explicit which assumptions reduce the full selection-response problem to the familiar two-parameter form \citep{1984MNRAS.206..377E,2021A&A...653A...9S}.
This perspective is useful, because the classical coefficient is still widely used in observational analyses, even when real catalogs deviate substantially from those assumptions.

\subsection{General response formula}

From Section~\ref{sec:general_formalism}, the first-order fluctuation in the number counts is
\begin{align}
    \frac{n(\nhat)-n_0(\nhat)}{n_0(\nhat)}
    =
    \mathcal A(\nhat)\,\bvec\cdot\nhat
    +
    \mathcal O(\beta^2),
\end{align}
where
\begin{align}
    \mathcal A(\nhat)
    =
    2+\frac{R(\nhat)}{n_0(\nhat)}.
\end{align}
The term $2$ is the universal contribution from aberration; $R(\nhat)$ encodes the Doppler response of the source population and survey selection.

This expression cleanly separates a purely geometric effect from the catalog-dependent response. All subsequent approximations can be viewed as progressively simplifying $R(\nhat)$ until it reduces to a small set of effective numbers.

\subsection{Uniform sky response}

First assume the selection function is independent of direction,
\begin{align}
    \mathcal W(\bm y,\nhat)=\mathcal W(\bm y),
\end{align}
and the completeness is spatially uniform.
Then $n_0(\nhat)=\bar n_0$ and $\mathcal A(\nhat)=\mathcal A$, so the fractional fluctuation reduces to a pure dipole:
\begin{align}
    \frac{n(\nhat)-\bar n_0}{\bar n_0}
    =
    \mathcal A\,\bvec\cdot\nhat.
\end{align}

This assumption concerns only survey uniformity, not the validity of the Ellis--Baldwin coefficient itself.
In practice, masks and depth variations are handled at the estimator level, but they must be conceptually distinguished from the intrinsic response coefficient \citep{2009ApJ...692..887C,2013A&A...555A.117R,2023A&A...675A..72W}.

\subsection{Single flux-threshold selection}

Next suppose the catalog is defined by a deterministic threshold in one band:
\begin{align}
    \mathcal W(\bm y)=\Theta(S_{\rm b}-S_{\min}).
\end{align}
The response is then dominated by sources near the survey limit and depends on the local slope of the cumulative counts.

We define
\begin{align}
    x
    \equiv
    -\left.
    \frac{\pd\ln N(>S)}{\pd\ln S}
    \right|_{S_{\min}},
    \label{eq:x_definition}
\end{align}
using the conventional positive-sign convention of dipole studies \citep{1984MNRAS.206..377E,2015APh....61....1T,2021A&A...653A...9S}.

The dipole coefficient then becomes
\begin{align}
    \mathcal A
    =
    2+
    x
    \left.
    \frac{\pd\ln S_{\rm b}}{\pd\ln\delta}
    \right|_{\delta=1}.
    \label{eq:A_2_plus_xq}
\end{align}

This step already compresses potentially complicated number-count behavior into a single threshold-local slope.
If the counts curve significantly with flux, $x$ becomes threshold-dependent and no longer characterizes the full catalog by one constant.

\subsection{Narrow band and power-law spectrum}

To recover the classical monochromatic treatment, assume a sufficiently narrow band and a local power-law spectrum
\begin{align}
    S'_{\nu'}(\nu')\propto \nu'^{-\alpha}.
\end{align}
For unresolved point sources,
\begin{align}
    S_\nu(\nu)=\delta\,S'_{\nu'}(\nu/\delta),
\end{align}
which immediately gives
\begin{align}
    S_\nu(\nu)\propto \delta^{1+\alpha}.
\end{align}
Hence, in the narrow-band limit,
\begin{align}
    \left.
    \frac{\pd\ln S_{\rm b}}{\pd\ln\delta}
    \right|_{\delta=1}
    =
    1+\alpha.
    \label{eq:q_equals_1_plus_alpha}
\end{align}
Here \(S_{\rm b}\) coincides with the effective monochromatic flux density.

This is the step where the full diversity of source SEDs is replaced by a single effective spectral index.
For mixed populations or broad filters the approximation can be inaccurate, especially when different source classes dominate at different flux thresholds.

\subsection{Power-law counts}

Finally, assume the cumulative counts themselves obey a power law:
\begin{align}
    N(>S)\propto S^{-x}.
\end{align}
Then \(x\) is constant and the detailed structure of the selection boundary collapses into a single number.
Combining Eqs.~\eqref{eq:A_2_plus_xq} and~\eqref{eq:q_equals_1_plus_alpha} yields
\begin{align}
    \mathcal A
    =
    2+x(1+\alpha),
\end{align}
which is exactly the Ellis--Baldwin coefficient.
The corresponding kinematic dipole vector is
\begin{align}
    \vec d_{\rm kin}
    =
    \left[2+x(1+\alpha)\right]\bvec.
\end{align}

This limiting form remains elegant and useful because it isolates the essential scaling with count slope and spectral index.
Its continued popularity in observational work is therefore understandable, especially when only limited catalog information is available.

\subsection{Interpretation}

{The controlled-limit derivation clarifies which information is discarded in the classical approximation: source-to-source SED variation, curvature or threshold dependence of the counts slope, multiband selections, color and photometric-redshift cuts, and direction-dependent survey response.
Many of these effects have already appeared implicitly in the observational dipole literature through running slopes, flux-threshold tests, estimator comparisons, mask treatments, and survey-specific calibrations \citep{2015APh....61....1T, 2023A&A...675A..72W, 2024MNRAS.535L..49V, 2025PhRvD.111l3547V, 2025A&A...697A.112W}.}

The present formalism unifies these ingredients at the level of the theoretical response coefficient, while retaining the Ellis--Baldwin result as a well-defined limiting case.
In this sense, the classical expression is not replaced but embedded within a broader and more realistic framework.

Population mixing in radio source counts is itself well known.
The additional point clarified here is how such mixing propagates explicitly into the kinematic dipole coefficient through population fractions, threshold-local count slopes, spectral-index distributions, and their covariance structure.
The response formalism therefore provides a natural and survey-specific language for translating source-population complexity into dipole predictions.

{
\section{Application: Radio source counts with AGN and SFG contributions}\label{sec:mixed_population_application}}

Real radio catalogs are not drawn from a single homogeneous source population. 
Wide and deep continuum surveys typically contain at least two dominant components: active galactic nuclei (AGNs) and star-forming galaxies (SFGs). 
Their relative abundances vary with flux threshold, observing frequency, and survey depth \citep{2014MNRAS.440.2791V,2015MNRAS.452.1263P,2016A&ARv..24...13P,2017A&A...602A...6S,kono2021empirical}.
This situation makes radio source counts a natural application of the general formalism developed above.
Whereas the classical Ellis--Baldwin coefficient assumes a single count slope and a single spectral index, a mixed population requires an average over source classes.

\subsection{Two-component form}

Let the parent population be decomposed as
\begin{align}
    f(\bm y)=f_{\rm AGN}(\bm y)+f_{\rm SFG}(\bm y).
\end{align}
Assuming a direction-independent flux threshold and working to first order in $\beta$, each component contributes
\begin{align}
    n_i(\nhat)
    =
    n_{0,i}
    \left[
    1+\mathcal A_i\,\bvec\cdot\nhat
    \right]
    +
    \mathcal O(\beta^2),
\end{align}
with $i\in\{{\rm AGN,SFG}\}$.
The total catalog then has
\begin{align}
    \mathcal A_{\rm mix}
    =
    \frac{
    n_{0,{\rm AGN}}\mathcal A_{\rm AGN}
    +
    n_{0,{\rm SFG}}\mathcal A_{\rm SFG}
    }{
    n_{0,{\rm AGN}}+n_{0,{\rm SFG}}
    }.
    \label{eq:A_mix_basic}
\end{align}

If the cumulative counts of each component have local slope
\begin{align}
    x_i
    \equiv
    -
    \left.
    \frac{\pd\ln N_i(>S)}{\pd\ln S}
    \right|_{S_{\min}},
\end{align}
and the radio spectrum is approximated by
\begin{align}
    S_\nu\propto \nu^{-\alpha_i},
\end{align}
then
\begin{align}
    \mathcal A_i
    =
    2+x_i(1+\alpha_i),
\end{align}
so that
\begin{align}
    \mathcal A_{\rm mix}
    =
    2+
    \frac{
    n_{0,{\rm AGN}}x_{\rm AGN}(1+\alpha_{\rm AGN})
    +
    n_{0,{\rm SFG}}x_{\rm SFG}(1+\alpha_{\rm SFG})
    }{
    n_{0,{\rm AGN}}+n_{0,{\rm SFG}}
    }.
    \label{eq:A_mix_two_pop}
\end{align}
Writing
\begin{align}
    w_i
    \equiv
    \frac{n_{0,i}}{\sum_j n_{0,j}},
    \qquad
    \sum_i w_i=1,
\end{align}
this becomes
\begin{align}
    \mathcal A_{\rm mix}
    =
    2+\sum_i w_i x_i(1+\alpha_i).
    \label{eq:A_mix_compact}
\end{align}

\subsection{Mean-value form and covariance term}

Equation~\eqref{eq:A_mix_compact} shows that the relevant quantity is the weighted mean of \(x(1+\alpha)\).
Defining
\begin{align}
    \langle Q\rangle_w \equiv \sum_i w_i Q_i,
\end{align}
we obtain
\begin{align}
    \mathcal A_{\rm mix}
    =
    2+\langle x(1+\alpha)\rangle_w.
\end{align}
Using
\begin{align}
    \langle x\alpha\rangle_w
    =
    \langle x\rangle_w\langle\alpha\rangle_w
    +
    \cov_w(x,\alpha),
\end{align}
gives
\begin{align}
    \mathcal A_{\rm mix}
    =
    2+
    \langle x\rangle_w
    \left(1+\langle\alpha\rangle_w\right)
    +
    \cov_w(x,\alpha).
    \label{eq:A_covariance}
\end{align}
Hence, if \(x\) and \(\alpha\) are effectively uncorrelated near the threshold,
\begin{align}
    \cov_w(x,\alpha)\simeq0,
\end{align}
then
\begin{align}
    \mathcal A_{\rm mix}
    \simeq
    2+\bar{x}_w(1+\bar{\alpha}_w).
\end{align}
The formalism therefore makes explicit why the simple mean-value replacement works when the covariance term is negligible.

\subsection{Worked example from 3\,GHz source counts}

As an illustrative example, we adopt representative values motivated by the VLA-COSMOS 3\,GHz sample analyzed by \citet{kono2021empirical}.
That catalog contains 5410 SFGs and 1908 AGNs and reports mean radio spectral indices
\begin{align}
    \langle\alpha_{\rm SFG}\rangle
    =
    0.87\pm0.15,
    \qquad
    \langle\alpha_{\rm AGN}\rangle
    =
    0.85\pm0.26,
\end{align}
under the convention \(S_\nu\propto\nu^{-\alpha}\).
For demonstration we adopt threshold-local cumulative slopes
\begin{align}
    x_{\rm SFG}=0.9,
    \qquad
    x_{\rm AGN}=1.2,
\end{align}
together with
\begin{align}
    \alpha_{\rm SFG}=0.87,
    \qquad
    \alpha_{\rm AGN}=0.85.
\end{align}
If \(f_{\rm AGN}\) denotes the AGN fraction, then
\begin{align}
    \mathcal A_{\rm mix}
    =
    2
    +
    f_{\rm AGN}x_{\rm AGN}(1+\alpha_{\rm AGN})
    +
    (1-f_{\rm AGN})x_{\rm SFG}(1+\alpha_{\rm SFG}),
\end{align}
or numerically,
\begin{align}
    \mathcal A_{\rm mix}
    =
    3.683+0.537\,f_{\rm AGN}.
    \label{eq:A_numeric_linear}
\end{align}
Thus
\begin{align}
    f_{\rm AGN}=0.2
    &\quad\Rightarrow\quad
    \mathcal A_{\rm mix}=3.790,\\[2pt]
    f_{\rm AGN}=0.5
    &\quad\Rightarrow\quad
    \mathcal A_{\rm mix}=3.952,\\[2pt]
    f_{\rm AGN}=0.8
    &\quad\Rightarrow\quad
    \mathcal A_{\rm mix}=4.113.
\end{align}
Using the CMB dipole speed \(\beta=1.233\times10^{-3}\), the corresponding dipole amplitudes are
\begin{align}
    d=\beta\mathcal A_{\rm mix}
    =
    (4.67,\ 4.87,\ 5.07)\times10^{-3}.
\end{align}
Even this simple two-component model produces several-percent shifts in the expected radio dipole amplitude as the source mixture changes.

\subsection{Implications}

This example shows that a single universal pair \((x,\alpha)\) is generally insufficient for mixed radio catalogs.
Flux-threshold dependence can arise not only from curvature in the counts but also from changes in the AGN/SFG mixture.
Precision comparisons between radio dipoles and the CMB kinematic dipole should therefore propagate uncertainties in source classification, spectral-index distributions, and population fractions.
In this sense, the Ellis--Baldwin coefficient is better regarded as a population-dependent response than as a fixed constant.

\section{Discussion and conclusions}
\label{sec:discussion}

\subsection{From theoretical response to observed dipoles}
\label{subsec:estimator_direction_amplitude}

The first-order prediction derived in this work is
\begin{align}
    \frac{n(\nhat)-n_0(\nhat)}{n_0(\nhat)}
    =
    \mathcal A(\nhat)\,\bvec\cdot\nhat
    +
    \mathcal O(\beta^2),
    \label{eq:frac_fluct_model}
\end{align}
where \(\mathcal A(\nhat)\) is the kinematic response coefficient.
What a catalog actually measures, however, is not \(\mathcal A\) itself but the dipole vector of the observed sky distribution.
In practice one must therefore separate the theoretical kinematic contribution from additional survey- and volume-dependent terms:
\begin{align}
    \vec d_{\rm est}
    =
    \vec d_{\rm kin}
    +
    \vec d_{\rm LSS}
    +
    \vec d_{\rm sys},
\end{align}
where \(\vec d_{\rm LSS}\) is the contribution from large-scale structure within the survey volume and \(\vec d_{\rm sys}\) collects observational systematics such as calibration errors, incompleteness, or stellar contamination \citep{2012MNRAS.427.1994G,2021Univ....7..107S,2025A&A...697A.112W}.
The framework developed here predicts
\begin{align}
    \vec d_{\rm kin}
    =
    \mathcal A\,\bvec
\end{align}
to first order, while the separation of \(\vec d_{\rm LSS}\) and \(\vec d_{\rm sys}\) remains survey-dependent.

This distinction is essential.
The dipole vector is the direct observable from data, whereas converting its amplitude to a velocity requires an independent evaluation of the survey-specific coefficient \(\mathcal A[\mathcal W,f]\).
Estimator-side uncertainty and theory-side response are therefore logically separate parts of the problem.

\subsubsection{Dipole-field model}

Define the fractional fluctuation field as
\begin{align}
    \Delta(\nhat)
    \equiv
    \frac{n(\nhat)-\bar n}{\bar n},
\end{align}
where \(\bar n\) is the all-sky mean (or the mean under the survey mask).
Its dipole component can be written
\begin{align}
    \Delta(\nhat)\simeq \vec d\cdot\nhat,
    \label{eq:dipole_linear_field}
\end{align}
with
\begin{align}
    \vec d=d\,\hat{\vec d},
    \qquad
    d\equiv|\vec d|,
    \qquad
    \hat{\vec d}\equiv \vec d/|\vec d|.
\end{align}
For a purely kinematic signal one expects
\begin{align}
    \vec d_{\rm kin}\simeq \mathcal A\,\bvec.
    \label{eq:d_kin_predict}
\end{align}

\subsubsection{Least-squares estimator}

For a catalog of \(N\) sources with directions \(\{\nhat_i\}_{i=1}^N\), one can fit the fluctuation field with the dipole model
\begin{align}
    \Delta_{\rm obs}(\nhat)\simeq \vec d\cdot\nhat.
\end{align}
After pixelization, a weighted least-squares estimator is obtained by minimizing
\begin{align}
    \chi^2(\vec d)
    \equiv
    \sum_p
    \omega_p
    \left[
    \Delta_p-\vec d\cdot\nhat_p
    \right]^2,
    \label{eq:chi2}
\end{align}
where \(p\) labels pixels, \(\Delta_p\) is the measured fluctuation, and \(\omega_p\) is a weight (typically proportional to \(N_p\) in the Poisson limit).
The normal equations are
\begin{align}
    \sum_j M_{ij}d_j &= h_i,
    \\
    M_{ij} &\equiv \sum_p \omega_p\,\hat n_{p,i}\hat n_{p,j},
    \\
    h_i &\equiv \sum_p \omega_p\,\Delta_p\,\hat n_{p,i},
\end{align}
with solution
\begin{align}
    \vec d_{\rm est}=\mathsf M^{-1}\vec h.
    \label{eq:d_est}
\end{align}
The estimated direction and amplitude are
\begin{align}
    \hat{\vec d}_{\rm est}
    =
    \frac{\vec d_{\rm est}}{|\vec d_{\rm est}|},
    \qquad
    d_{\rm est}=|\vec d_{\rm est}|.
    \label{eq:dir_amp}
\end{align}
Under Poisson-distributed counts this estimator is equivalent to maximum likelihood; the derivation is given in Appendix~\ref{app:poisson_likelihood} and closely related forms have been used in previous work \citep{2009ApJ...692..887C,2013A&A...555A.117R,2015APh....61....1T}.

Beyond linear and quadratic estimators, Bayesian approaches have also been developed for cosmic-dipole analyses.
These methods infer the posterior distribution of the dipole amplitude and direction conditioned on the observed catalog, naturally propagating shot noise, incomplete sky coverage, and nuisance contributions such as calibration uncertainties or residual large-scale-structure contamination.
Recent implementations range from Bayesian estimators for wide-area radio surveys to likelihood analyses that explicitly incorporate overdispersed source-count statistics \citep{2023A&A...675A..72W, 2024MNRAS.535L..49V, 2025PhRvL.135t1001B, 2025PhRvD.111l3547V}.
Within the present formalism, \(\mathcal A[\mathcal W,f]\) may be treated either as a deterministic theoretical input or, more generally, as a hyper-parameter with prior uncertainty reflecting imperfect knowledge of the source population and selection model.

\subsubsection{From dipoles to velocities}

If \(\mathcal A\) is known, the estimated dipole can be mapped to an effective velocity:
\begin{align}
    \vec d_{\rm est}\simeq \mathcal A\,\bvec,
\end{align}
so that
\begin{align}
    \bvec_{\rm est}
    \simeq
    \frac{\vec d_{\rm est}}{\mathcal A},
    \qquad
    \beta_{\rm est}
    =
    \frac{d_{\rm est}}{\mathcal A},
    \qquad
    \hat{\bvec}_{\rm est}
    =
    \hat{\vec d}_{\rm est}.
    \label{eq:beta_from_d}
\end{align}
Conversely, if the CMB dipole velocity \(\bvec_{\rm CMB}\) is adopted as reference, the predicted source-count dipole is
\begin{align}
    d_{\rm pred}
    =
    \mathcal A\,|\bvec_{\rm CMB}|,
    \qquad
    \hat{\vec d}_{\rm pred}
    =
    \hat{\bvec}_{\rm CMB}.
\end{align}
Observed and predicted dipoles can then be compared through the amplitude ratio \(|\vec d_{\rm est}|/d_{\rm pred}\) and the angular separation \(\cos^{-1}(\hat{\vec d}_{\rm est}\cdot\hat{\bvec}_{\rm CMB})\).

\subsubsection{Background model and practical evaluation of \texorpdfstring{$\mathcal A$}{A}}
\label{subsec:A_practical}

In practice, constructing \(\Delta_p\) requires a background model for the expected counts.
A common factorized form is
\begin{align}
    \lambda(\nhat)=\bar n\,W(\nhat)\,C(\nhat),
\end{align}
where \(\lambda(\nhat)\) is the expected background count density, \(W\) is the geometric mask, and \(C\) denotes effective completeness (including depth variations and contamination corrections).

One then uses
\begin{align}
    \Delta_p=\frac{N_p-\lambda_p}{\lambda_p}.
\end{align}
Within the present formalism these ingredients are absorbed into a single effective selection function \(\mathcal W(\bm y,\nhat)\).
At the level of dipole estimation the practical task is straightforward: construct the fluctuation field relative to the expected background and fit its dipole component.

Theoretical interpretation, however, requires a separate evaluation of \(\mathcal A\).
In the general framework,
\begin{align}
    \mathcal A
    =
    2+\frac{R}{n_0},
\end{align}
or equivalently equation~\eqref{eq:A_general}.
Thus \(\mathcal A\) is not a universal constant but a functional of the catalog selection and source population.
In practice it can be obtained by numerical integration from a model for \(f(\bm y)\), by the discrete estimator of Section~\ref{subsec:discrete_sum}, or by semi-analytic approximations using SED templates and gradients of the selection function.
The dipole direction is therefore determined by the observed sky distribution, whereas the physical interpretation of the amplitude depends directly on how accurately \(\mathcal A\) is computed.

\subsection{Interpreting tensions in observed dipole measurements}

Measurements of number-count dipoles in radio, infrared, and optical catalogs are often broadly aligned with the CMB dipole, but reported amplitudes and sometimes directions show non-negligible survey-to-survey variation.
As emphasized by \citet{2012MNRAS.427.1994G}, such discrepancies do not by themselves imply a failure of the kinematic interpretation.
They may arise from local-structure contamination, survey systematics, estimator choices, or from an incomplete treatment of the survey-dependent response coefficient.
This point is illustrated both by mid-infrared AGN dipoles \citep{2021Univ....7..107S} and by recent multi-survey analyses \citep{2025A&A...697A.112W}.

The present formalism sharpens this distinction.
If different catalogs are analyzed with the same simplified coefficient \(2+x(1+\alpha)\) even though their selections, bandpasses, SED distributions, and source mixtures differ, then differences in the predicted kinematic amplitude are absorbed incorrectly into the observational residual.
Apparent tension may then reflect a mismatch between the theoretical response model and the actual catalog construction rather than a genuine inconsistency with the CMB dipole.

This does not eliminate the need to model \(\vec d_{\rm LSS}\) and \(\vec d_{\rm sys}\).
On the contrary, the separation
\begin{align}
    \vec d_{\rm est}
    =
    \vec d_{\rm kin}
    +
    \vec d_{\rm LSS}
    +
    \vec d_{\rm sys}
\end{align}
makes clear that estimator uncertainties, large-scale-structure contamination, and theoretical response should be treated as distinct components of the inference problem.
Bayesian hierarchical approaches are especially attractive in this context because they allow nuisance contributions to be marginalized jointly with the kinematic signal.

\subsection{Summary and outlook}
\label{subsec:conclusion_outlook}

We have developed a general theoretical framework for the kinematic dipole in source number counts that does not assume power-law counts, a single spectral index, or a monochromatic flux limit.
For a general parent population \(f(\bm y)\) and a general selection function \(\mathcal W(\bm y,\nhat)\), including finite bandpasses, multiband photometry, color and photometric-redshift cuts, weights, and direction-dependent survey depth, the dipole coefficient is given by the functional response
\begin{align}
    \mathcal A(\nhat)
    =
    2+
    \frac{R(\nhat)}{n_0(\nhat)}.
\end{align}

This identifies the kinematic dipole amplitude as the Doppler response of the catalog construction itself.

We have shown that the classical Ellis--Baldwin result is recovered as a controlled limiting case, and that realistic source mixtures such as AGN+SFG radio catalogs naturally lead to population-averaged response coefficients with covariance corrections.
We have also clarified the division of roles between dipole estimators, which operate directly on the sky distribution, and the theoretical coefficient \(\mathcal A[\mathcal W,f]\), which translates an observed dipole into a physical velocity.

Several next steps are especially important.
Survey-specific selection functions should be made explicit at the level of catalog pipelines and used to evaluate \(\mathcal A[\mathcal W,f]\) numerically.
Large-scale-structure and systematic contributions should be quantified with mocks, tomographic analyses, and null tests.
Finally, future wide-area surveys such as LSST, Euclid, and SKA will benefit from standardized representations of direction-dependent depth and probabilistic completeness, making survey-specific dipole predictions directly comparable.

Once these ingredients are treated consistently, comparisons with the CMB dipole can move beyond a universal coefficient and toward survey-specific theoretical predictions.
Only at that stage can residual discrepancies be interpreted robustly in terms of astrophysical contamination, observational systematics, or genuinely new physics.

\begin{acknowledgements}
{We thank the referee for constructive comments and suggestions that significantly improved the clarity of this paper.}
This work was supported by the JSPS Grant-in-Aid for Scientific Research (24H00247), and by the Joint Research Program (General Research~2) of the Institute of Statistical Mathematics,  ``Machine-Learning-Based Cosmogony: From Structure Formation to Galaxy Evolution''. 
\end{acknowledgements}

\appendix

\section{Equivalence between Poisson-likelihood and least-squares dipole estimation}
\label{app:poisson_likelihood}

This appendix shows that the least-squares dipole estimator used in Section~\ref{subsec:estimator_direction_amplitude} follows naturally from likelihood maximization for Poisson-distributed counts.
Related discussions of dipole detectability and estimator properties may be found in the literature \citep[e.g.][]{2009ApJ...692..887C,2013A&A...555A.117R}.

\subsection{Poisson model for pixelized counts}

Divide the sky into sufficiently fine pixels and let $N_p$ be the observed count in pixel $p$.
Under the dipole model, the expectation value is
\begin{align}
	\lambda_p(\vec d)
	\equiv
	\bar n\,W(\hat{\vec n}_p)
	\left[
	1+\vec d\cdot\hat{\vec n}_p
	\right],
	\label{eq:lambda_p_app}
\end{align}
where $\bar n$ is the mean surface density and $W(\hat{\vec n})$ is the observational window function including the mask and effective completeness.
Since we work only to first order in the dipole, $\vec d\cdot\hat{\vec n}_p$ is treated as a small perturbation.

Assuming independent Poisson counts, the likelihood is
\begin{align}
	\mathcal L(\vec d)
	=
	\prod_p
	\frac{\lambda_p(\vec d)^{N_p}}{N_p!}
	\exp\!\left[-\lambda_p(\vec d)\right],
\end{align}
with log-likelihood
\begin{align}
	\ln\mathcal L(\vec d)
	=
	\sum_p
	\left[
	N_p\ln\lambda_p(\vec d)
	-\lambda_p(\vec d)
	-\ln N_p!
	\right].
	\label{eq:loglike_app}
\end{align}

\subsection{Normal equations from likelihood maximization}

Expanding $\ln\lambda_p(\vec d)$ to first order in $\vec d$ gives
\begin{align}
	\ln\lambda_p
	=
	\ln\!\left[\bar n\,W(\hat{\vec n}_p)\right]
	+
	\vec d\cdot\hat{\vec n}_p
	+
	\mathcal O(d^2).
\end{align}
Substituting this into equation~\eqref{eq:loglike_app} and imposing $\partial\ln\mathcal L/\partial d_i=0$ yields
\begin{align}
	\frac{\partial\ln\mathcal L}{\partial d_i}
	=
	\sum_p
	\left[
	N_p-\lambda_p(\vec 0)
	\right]
	\hat n_{p,i}
	+
	\mathcal O(d)
	=
	0,
\end{align}
where $\lambda_p(\vec 0)=\bar n\,W(\hat{\vec n}_p)$.
To first order, this reduces to
\begin{align}
	\sum_j M_{ij}\,d_j = h_i,
	\label{eq:normal_app}
\end{align}
with
\begin{align}
	M_{ij}
	&\equiv
	\sum_p
	\bar n\,W(\hat{\vec n}_p)\,
	\hat n_{p,i}\hat n_{p,j},
	\\
	h_i
	&\equiv
	\sum_p
	\left[
	N_p-\bar n\,W(\hat{\vec n}_p)
	\right]
	\hat n_{p,i}.
\end{align}
Hence
\begin{align}
	\vec d_{\rm est}
	=
	\mathsf M^{-1}\vec h.
\end{align}

\subsection{Relation to weighted least squares}

Using the Gaussian approximation to the Poisson likelihood,
\begin{align}
	-2\ln\mathcal L
	\simeq
	\sum_p
	\frac{(N_p-\lambda_p)^2}{\lambda_p}
	+
	\mathrm{const.},
\end{align}
which is equivalent to weighted least squares with variance $\sigma_p^2=\lambda_p$.
Accordingly, the estimator in the main text,
\begin{align}
	\chi^2
	=
	\sum_p
	\omega_p
	\left[
	\Delta_p-\vec d\cdot\hat{\vec n}_p
	\right]^2,
	\qquad
	\omega_p\propto\lambda_p,
\end{align}
is the natural quadratic approximation to Poisson-likelihood maximization.
The same likelihood may also be embedded in a Bayesian analysis by combining it with priors for the dipole vector and nuisance parameters.

\section{Dipole amplitude as a functional and statistical uncertainty}
\label{app:fisher_covariance}

The central result of this paper is that the kinematic dipole amplitude is a functional
\begin{align}
    \mathcal A
    \equiv
    \mathcal A[\mathcal W,f],
\end{align}
depending on the selection function $\mathcal W$ and the parent population $f(\bm y)$, and the family of source SEDs.
This appendix summarizes how the estimated dipole vector acquires statistical uncertainty for finite samples and incomplete sky coverage, given a specified response coefficient.

\subsection{Theoretical input and observational uncertainty}

In the main text, the first-order fluctuation field is
\begin{align}
	\Delta(\nhat)
	\equiv
	\frac{n(\nhat)-n_0(\nhat)}{n_0(\nhat)}
	\simeq
	\mathcal A(\nhat)\,\bvec\cdot\nhat.
\end{align}
In the uniform-sky limit, $\mathcal A(\nhat)\to\mathcal A$, so the theoretical prediction becomes
\begin{align}
	\vec d_{\rm kin}
	\equiv
	\mathcal A\,\bvec.
\end{align}
In what follows, $\vec d_{\rm kin}$ is treated as a known theoretical input.
The statistical discussion is therefore separate from the construction of the functional $\mathcal A$ itself.

\subsection{Fisher matrix for the dipole estimator}

Using the same pixelized Poisson model as in Appendix~\ref{app:poisson_likelihood},
\begin{align}
	\lambda_p(\vec d)
	\equiv
	\bar n\,W(\hat{\vec n}_p)
	\left[
	1+\vec d\cdot\hat{\vec n}_p
	\right],
\end{align}
the Fisher matrix is
\begin{align}
	F_{ij}
	=
	\sum_p
	\bar n\,W(\hat{\vec n}_p)\,
	\hat n_{p,i}\hat n_{p,j}.
	\label{eq:Fisher_general}
\end{align}
This is the same matrix that appears in the normal equations of the least-squares estimator.
By the Cram\'er--Rao bound,
\begin{align}
	\cov (d_i,d_j)
	=
	(F^{-1})_{ij}.
\end{align}

The roles of theory and statistics are thus distinct.
The Fisher matrix is determined by sample size and survey geometry, whereas the functional $\mathcal A$ sets the mean signal,
\begin{align}
	\vec d_{\rm est}
	=
	\mathcal A\,\bvec
	+
	\text{(statistical fluctuations)}.
\end{align}
The new ingredient introduced in this paper is the explicit form of the first term.

\subsection{Classical limit}

In the full-sky and uniform-completeness limit,
\begin{align}
	F_{ij}
	=
	\frac{\bar N}{3}\,\delta_{ij},
	\qquad
	\cov (d_i,d_j)
	=
	\frac{3}{\bar N}\,\delta_{ij}.
\end{align}
This statistical structure is standard and is shared by classical Ellis--Baldwin analyses.
What changes in the present framework is not the Fisher matrix itself, but the theoretical response coefficient placed on top of it.
In the classical approximation, $\mathcal A=2+x(1+\alpha)$ is a constant; in the general case, $\mathcal A$ depends on the detailed selection and source population.

Even in the absence of a true dipole, finite sampling produces a noise dipole in $\vec d_{\rm est}$.
In the full-sky limit, the amplitude $d=|\vec d_{\rm est}|$ follows a Maxwell distribution, which can be used to assess the statistical significance of a measured signal.
The purpose of this appendix is therefore not to introduce a new estimator, but to clarify how the functional response coefficient of the main text interfaces with standard estimator uncertainties.

\bibliographystyle{aasjournalv7}
\bibliography{cosmic_dipole}

\end{document}